\documentclass[%
 reprint,
nofootinbib,
 amsmath,amssymb,
 aps,
]{revtex4-1}

\usepackage{color}
\usepackage{graphicx}
\usepackage{dcolumn}
\usepackage{bm}
\usepackage[utf8]{inputenc} 

\begin{document}

\preprint{APS/123-QED}

\title{Compactness in the Thermal Evolution of Twin Stars}

\author{F. Lyra, L. Moreira}
\affiliation{Instituto de Fisica, Universidade Federal Fluminense, Niteroi, Brazil}

\author{R. Negreiros}
\affiliation{Instituto de Fisica, Universidade Federal Fluminense, Niteroi, Brazil}

\author{R. O. Gomes}
\affiliation{Frankfurt Institute for Advanced Studies (FIAS), Frankfurt, Germany}

\author{V. Dexheimer}
\affiliation{Department of Physics, Kent State University, Kent OH 44242 USA}

\date{\today}

\begin{abstract}
In this work, we study for the first time the thermal evolution of twin star pairs, i.e., stars that present the same mass but different radius and compactness. We collect available equations of state that give origin to a second branch of stable compact stars with quarks in their core. For each equation of state, we investigate the particle composition inside stars and how differently each twin evolves over time, which depends on the central density/pressure and consequent crossing of the threshold for the Urca cooling process. We find that, although the general stellar thermal evolution depends on mass and particle composition, withing one equation of state, only twin pairs that differ considerably on compactness can be clearly distinguished by how they cool down.

\end{abstract}

\pacs{Valid PACS appear here}
\maketitle


\section{Introduction}

There are many different observables that can be used to probe the interior of neutron stars. Unfortunately, common properties such as stellar masses and radii are often not enough to distinguish between stars that have a small deconfined quark matter core in their interiors \cite{Alford:2004pf}. One exception, are stars that differ considerably (observation wise) from their original-branch counterparts, creating the so-called twin-star configuration in the mass-radius diagram. Since the original idea was discussed in 1968 \cite{Gerlach:1968}, many works discussing this configuration appeared in the literature including, for example, studies of how they could be enhanced or suppressed by strong magnetic fields \cite{Gomes:2018bpw} and rotation \cite{Bhattacharyya:2004fn,Banik:2004ju,Bejger:2016emu,Bozzola:2019tit}, and how they can leave distinguishable imprints in gravitational-wave signals form compact-star mergers \cite{Weih:2019xvw,Montana:2018bkb,Tan:2021nat}.

The main idea behind twins is that when the central stellar density is above the threshold for a given strong phase transition (where the speed of sound is zero or low for an extended region of density) the sudden softness of the equation of state (EoS) destabilizes stars. On the other hand, if the central stellar density is large enough for a sizable core that contains this new phase, which must necessarily be described by a very stiff EoS, it can turn the second branch of the stellar sequence stable. This can happen, for example, through a first order phase transition from hadronic to quark matter at stellar cores. Studies also show that a third branch of stars can arise from a second transition to CFL matter \cite{Banik:2002kc,Alford:2017qgh}. But, more importantly, this new configuration is much smaller. In this way, there can be two twin stars with the same mass, the original one and the small, more compact one due to them having different particles content. 

In recent years, the observation of neutron stars has achieved remarkable advances, such as the new results from NICER \cite{Miller:2019cac,Riley:2019yda,Riley:2021pdl,Miller:2021qha} and the detection of binary neutron stars mergers at LIGO and VIRGO \cite{TheLIGOScientific:2017qsa,Yagi:2016bkt}. Both electromagnetic and gravitational observational methods seek finding, in particular, a more accurate way to determine neutron star radii. Equivalent efforts were made in the past to successfully measure the mass of neutron stars. Looking for different observables to complementary probe the composition and structure of neutron stars is crucial in order to obtain a complete understanding of nuclear matter's behavior in the high-density regime. One alternative to the observables mentioned above is the thermal evolution of neutron stars. 

In this work, we explore the intriguing possibility of twin star configurations under the context of thermal evolution. The cooling of neutron stars is strongly dependent on both micro and macroscopic properties \cite{TSURUTA1965,Maxwell1979,Page2006,Weber2007,Negreiros2010}, thus serving as a powerful tool for probing the composition and structure of neutron stars in the context of several phenomena \cite{Horvath1991,Blaschke2000a,Shovkovy2002,Alford2005a, Negreiros2012,Negreiros2017,Levenfish1994,Page2009,Fortin:2017rxq,Aguilera2008,Pons2009, Weber2005,Alford2005}. This study follows up on an investigation done in Ref.~\cite{PhysRevC.91.055808}, in which the role of strangeness on the thermal evolution of neutron stars was investigated. Twin star configurations were identified in this previous work, but since the scope of that project was to study strangeness,  their thermal properties were not investigated. 

Here, we continue such study, focusing on the properties of twin star pairs, i.e., the hybrid stars in the second branch with respect to their original-branch counterparts. We look for possible thermal signatures that may allow us to identify twin stars via their thermal behavior (and compactness). In order to deepen our analysis, we have extended our study to a second model that also reproduces twin stars, although with different microscopic properties. With this, we aim to identify trends and differences in the thermal behavior of twin stars. To achieve such goal, we perform a thorough investigation of the cooling of twin stars pairs, taking into account all possible thermal processes and different setups for pairing in booth phases. To the extent of our knowledge, this is the first work in which the cooling of twin-star pairs is investigated. We consider this to be a first step in better understanding the properties of this intriguing possibility, which might be the first solid evidence for a strong phase transition to stable quark matter in our universe.

The manuscript is organized as follow: in section II, we discuss the relativistic mean field models used to describe a twin-star configuration and their particle populations; in section III, the macroscopic structure of stars is determined; in section IV, we present the formalism used for their thermal evolution; and, finally, in section V we present a discussion and draw our conclusions. 

\section{Equations of State}

\begin{figure}
\vspace{3mm}
\includegraphics[width=11.9cm,clip,trim=.63cm 4.4cm 0 3.21cm]{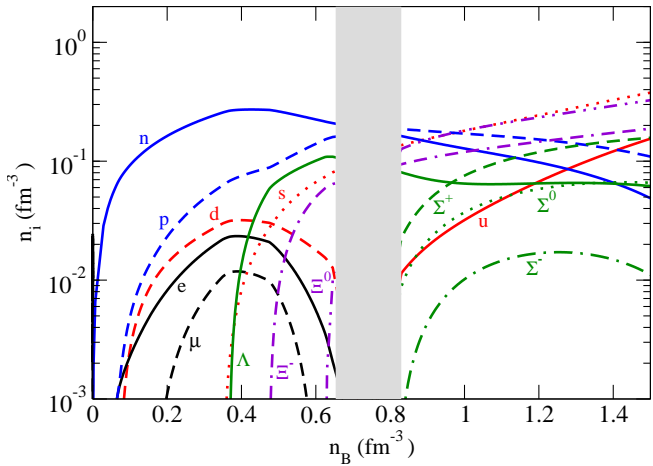}
\caption{Particle composition given by equation of state 1. The shaded region marks the baryon number density jump across the phase transition.
\label{part1}}
\end{figure}

In order to model the low-density portion of neutron stars corresponding to the crust, we use a separate EoS, which will be described in the next section. For the intermediate and high density parts of the EoS, which make up most of their radii, we make use of two distinct relativistic prescriptions described in the following.

\subsection{EoS 1: Excluded-Volume CMF model}

In this case, only one model is used to describe the intermediate and high density regions of neutron stars. The Chiral Mean Field (CMF) model, as described in Refs.~\cite{Dexheimer:2009hi,Hempel:2013tfa,Roark:2018uls,Roark:2018boj,Aryal:2020ocm,Dexheimer:2020rlp} contains the baryon octet and the three light quarks interacting through a mean field of mesons (plus free leptons). The field $\Phi$, which works as an order parameter for deconfinement in this formalism, modifies the mass of the fermions, suppressing the quark (or hadronic) degrees of freedom at low (or high) densities. While the hadronic part of the model has been fitted to reproduce nuclear and astrophysical properties, the quarks part has been fitted to reproduce lattice QCD data.

\begin{figure}
\vspace{3mm}
\includegraphics[width=11.9cm,clip,trim=.63cm 4.4cm 0 3.21cm]{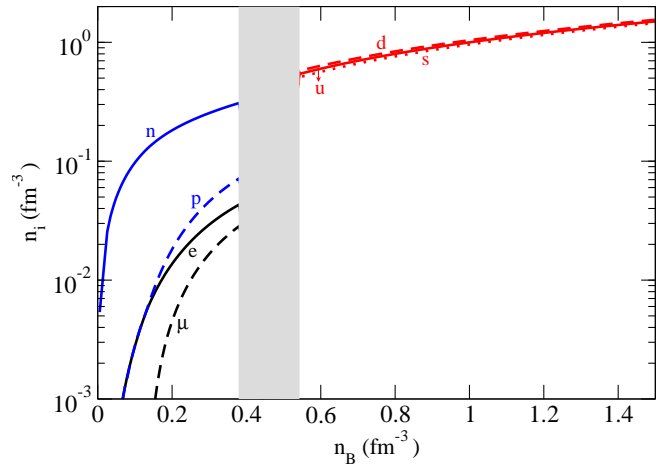}
\caption{Same as Fig.~\ref{part2} but for the equation of state 2.
\label{part2}}
\end{figure}

Here, we use a different version of the CMF model that also has a finite size for the baryons, included using the excluded volume prescription described in Refs.~\cite{PhysRevC.84.045208}. This version of the model has already been used to investigate the effects of strangeness content \cite{Dexheimer:2014pea} and strong magnetic fields \cite{Gomes:2018bpw} in neutron stars.\footnote{Note that an alternative version of the CMF model includes in addition the chiral partners of the baryons \cite{Dexheimer:2007tn,Dexheimer:2008cv,PhysRevC.87.015804,Mukherjee:2017jzi,Motornenko:2019arp}.} The appearance of deconfined quarks, which had become a smooth crossover (even at zero temperature) due to the excluded volume, can turn into a strong first-order phase transition for certain values of the quark coupling to the strange vector meson. This, in turn, can produce a twin configuration in which the compact twin star contains a very large of  strangeness fraction of $Y_S = 1.68$ (when compared to $Y_S = 0.01$ in the original branch). The strangeness fraction is defined as the sum of hadronic and quark strangeness normalized by the respective number densities $Y_S =\sum_i {S}_i n_i/\sum_i n_i$.

The particle population reproduced by this prescription is shown in Fig.~\ref{part1}, where the quark number densities were divided by $3$. The x-axis corresponds to the entire density regime inside the most massive compact twin reproduced by the model. Stars in the less compact branch of twins only reach densities to the left side of the shaded region, which marks the density jump created by the first-order phase transition. It can be seen that both phases contain hadrons and quarks, even though the low density phase is dominated by hadrons and the dense one by quarks.

\subsection{EoS 2: MBF\,+\,vBAG models}

\begin{table}
\begin{tabular}{|c|c|c|c|c|}
\hline         
EoS                     & \multicolumn{2}{c|}{EoS 1} & \multicolumn{2}{c|}{EoS 2} \\ \hline
branch                  & compact  & original        & compact  & original        \\ \hline
$M$ (M$_{\odot})$       & 1.68     & 1.68            & 1.97     & 1.97            \\ \hline
$R$ (km)                & 9.93     & 13.96           & 13.40    & 14.26           \\ \hline
$\tilde{\Lambda}$       & 9.39     & 324.68          & 66.38    & 122.84          \\ \hline
$C$                     & 0.26     & 0.18            & 0.22     & 0.20            \\ \hline
${n_B}_c$ (fm$^{-3})$   & 1.47     & 0.36            & 0.69     & 0.56            \\ \hline
$P_c$ (MeV/fm$^{3})$   & 839.39   & 54.20           & 158.95   & 81.80            \\\hline     
\end{tabular}
\caption{Characteristics of the most massive twin stars reproduced by both equations of state investigated, which include stellar mass, radius, tidal deformability, compactness $M/R$, central baryon number density, and pressure, in both the original and compact branches. }
\label{tabela}   
\end{table}

In this case, we combine two separate models connected through a first order phase-transition. For the hadronic phase, we select the Many-Body Forces (MBF) model including nucleons and leptons. In this relativistic mean-field framework, many-body force contributions are introduced through non-linear baryon couplings to the scalar fields mesons,  which are controlled by a parameter $\zeta$ \cite{Gomes:2014aka}. In this work, we choose $\zeta = 0.04$, which is the stiffest possible realistic parametrization of the model. For the quark phase, we select the MIT Bag model with vector interaction \cite{Franzon:2016urz}, which allows for a stiff EoS able to describe massive hybrid stars. Values of the vector coupling, bag constant and strange quark mass in the MIT bag model that give rise to a second branch are $(g_V /m_V)^2= 1.7$ fm$^2$, $B^{1/4} = 171$ MeV, and $m_S = 150$ MeV, respectively \cite{Gomes:2018eiv,Gomes:2018bpw}. The hadronic and quark EoSs are connected by a Maxwell construction, reproducing a sharp phase transition between the phases.

The particle population reproduced by this prescription is shown in Fig.~\ref{part2}. The quark number densities were once more divided by $3$. This is a much more traditional setup, in which the low density phase only contains hadrons (and leptons) and the dense one only contains quarks. The jump in baryon number density across the phase transition reproduced by EoS 2 is smaller than the one reproduced by EoS 1, but still comparable. Such behavior comes from a difference on the stiffness in the hadronic and quark phases withing the two prescriptions. Once more, the less compact stars of the twin pairs only reach densities to the left of the shaded region, while the more compact ones reach densities to the right.

The MBF model has been successfully applied to investigate a broad range of neutron stars topics, such as hyperonic stars \cite{Gomes:2014aka}, magnetic fields \cite{Dexheimer:2016yqu,Gomes:2017zkc,Gomes:2019paw} and tidal deformability \cite{Gomes:2018eiv,Dexheimer:2018dhb}. The vBAg model is implemented in a similar fashion as Refs.~\cite{Klahn:2017exz,Fischer:2016ojn,Klahn:2016uce,Cierniak:2019hhe}, and has been applied to describe hybrid and twin stars \cite{Gomes:2018eiv,Gomes:2018bpw,Jakobus:2020nxw,Dexheimer:2020rlp}. 

\section{Stellar Structure}

\begin{figure}
\vspace{3mm}
\includegraphics[width=8.9cm,clip,trim=.1cm 0 0 .77cm]{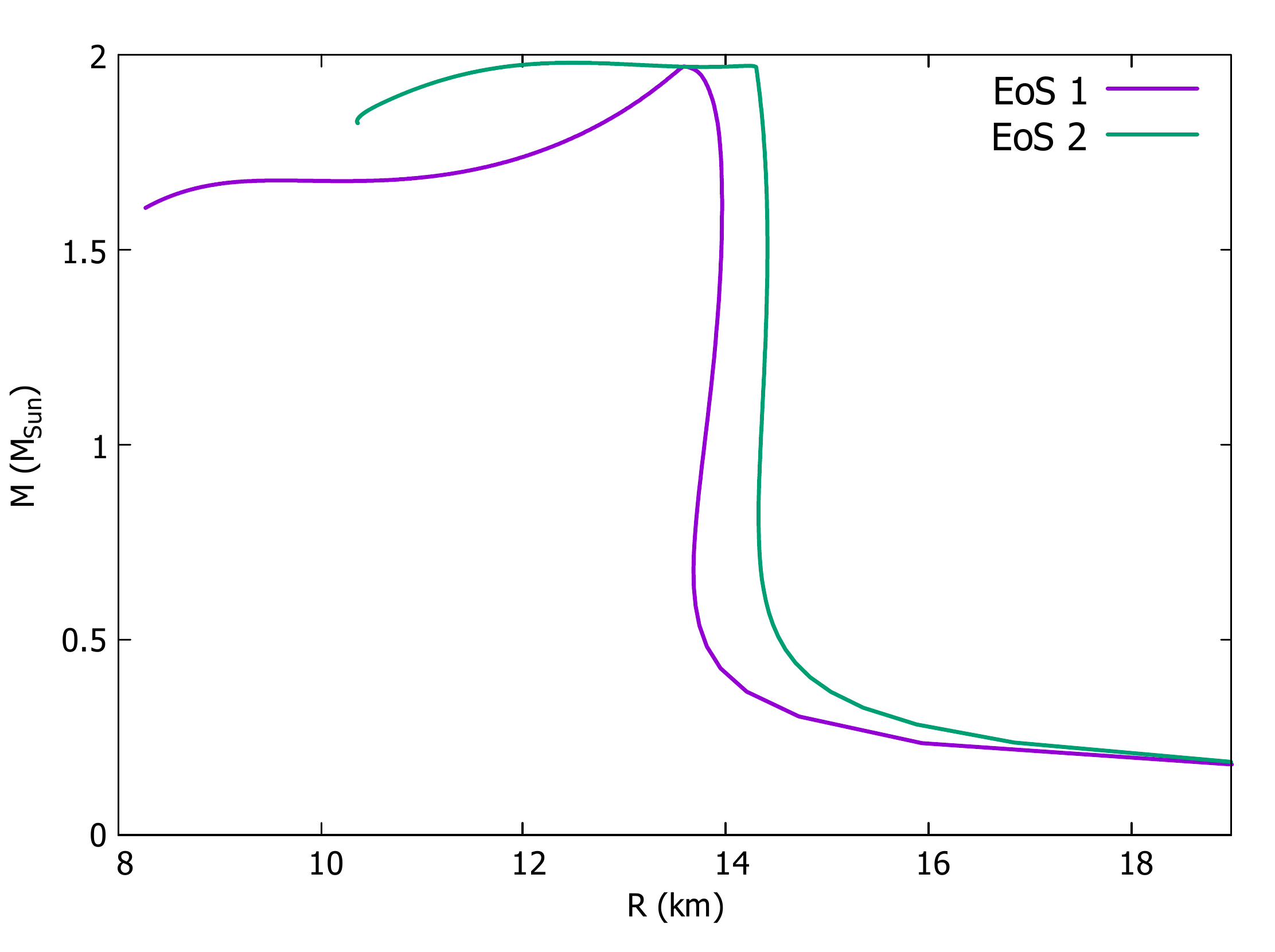}
\includegraphics[width=8.9cm,clip,trim=.1cm 0 0 .94]{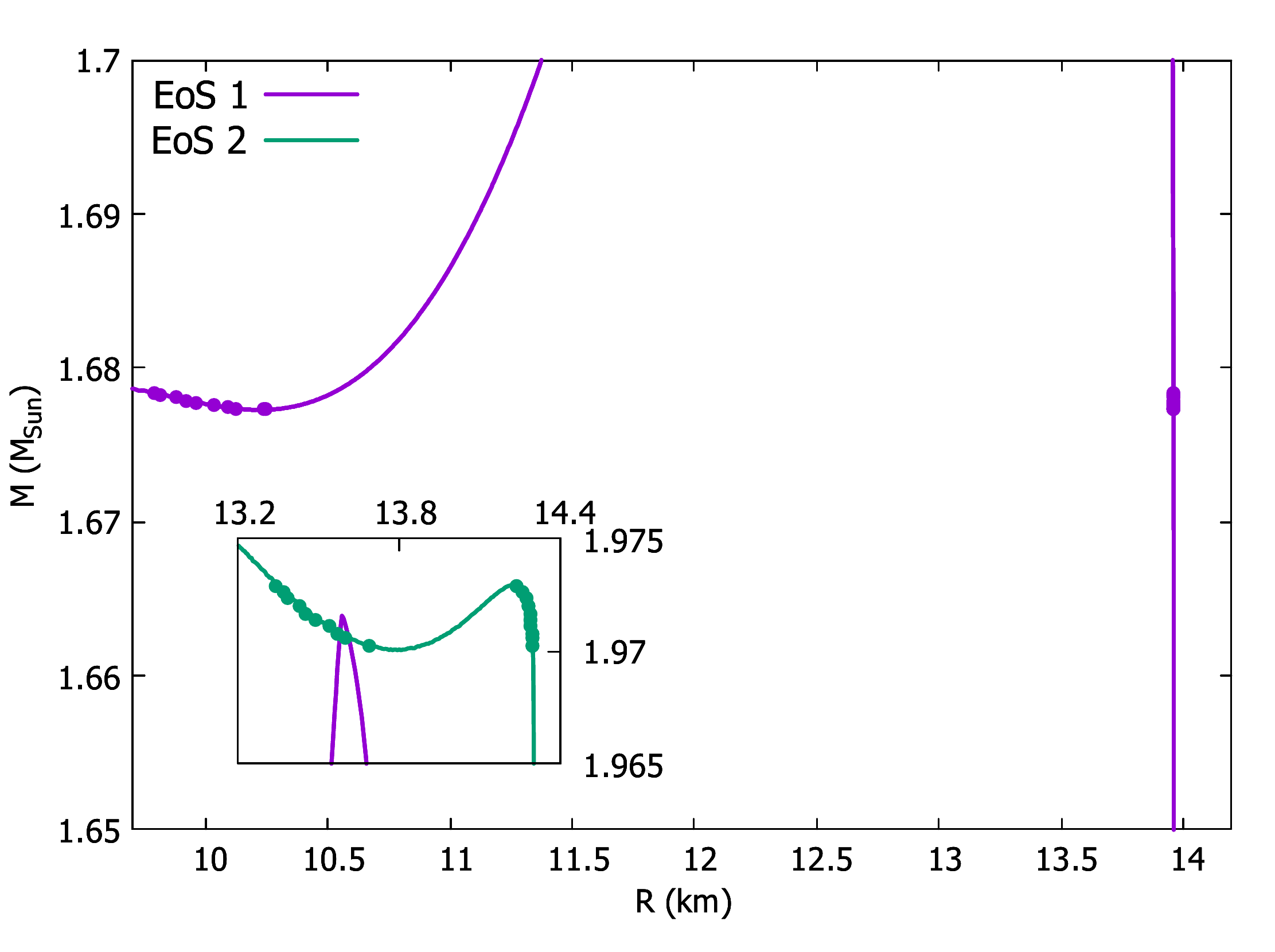}
\caption{Top panel: Mass-Radius diagram for both equations of state. Bottom panel: zoomed-in region from top panel with circles representing stable configurations in which pairs of stars with the same mass fulfill ${\partial M}/{\partial R} <0$.
\label{twin_ver}}
\end{figure}

Given our goal to calculate the thermal evolution of the twin stars generated by both of the EoSs presented in the last section, our next step is to calculate and analyze their macroscopic structures. In order to do so, we add the widely-used Baym-Pethick-Sutherland \cite{Baym:1971pw} EoS for the neutron-star crust. It describes the crust as composed of heavy ions sitting on a crystalline lattice, permeated by an electron gas, as well as a neutron gas for densities above that of the neutron drip. From the mass-radius  curve of each EoS calculated using the Tolman-Oppenheimer-Volkoff equations, it can be seen in the top panel of Fig.~\ref{twin_ver} that the mass region that contains the twin stars has both stable and unstable hydrostatic solutions. As the thermal evolution of neutron stars occurs over the period of several million years, we employ our efforts in the evaluation of the stable structures. We refer the reader to Ref.~\cite{Tonetto:2020bie} for the discussion of the unstable structures.

\begin{figure}
\vspace{3mm}
\includegraphics[width=11.9cm,clip,trim=1.21cm 3.5cm 0 3.21cm]{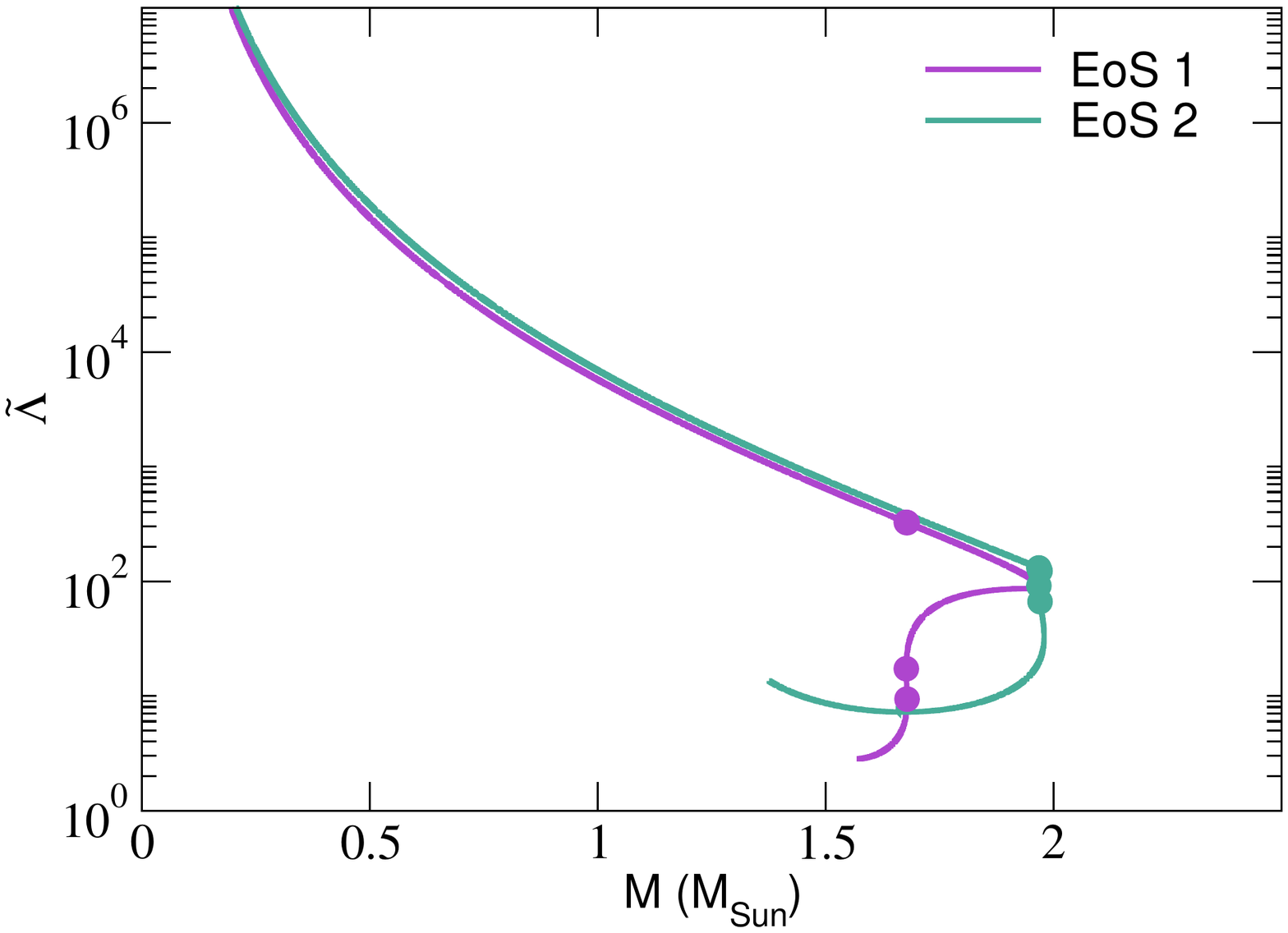}
\includegraphics[width=11.9cm,clip,trim=1.21cm 4.4cm 0 3.21cm]{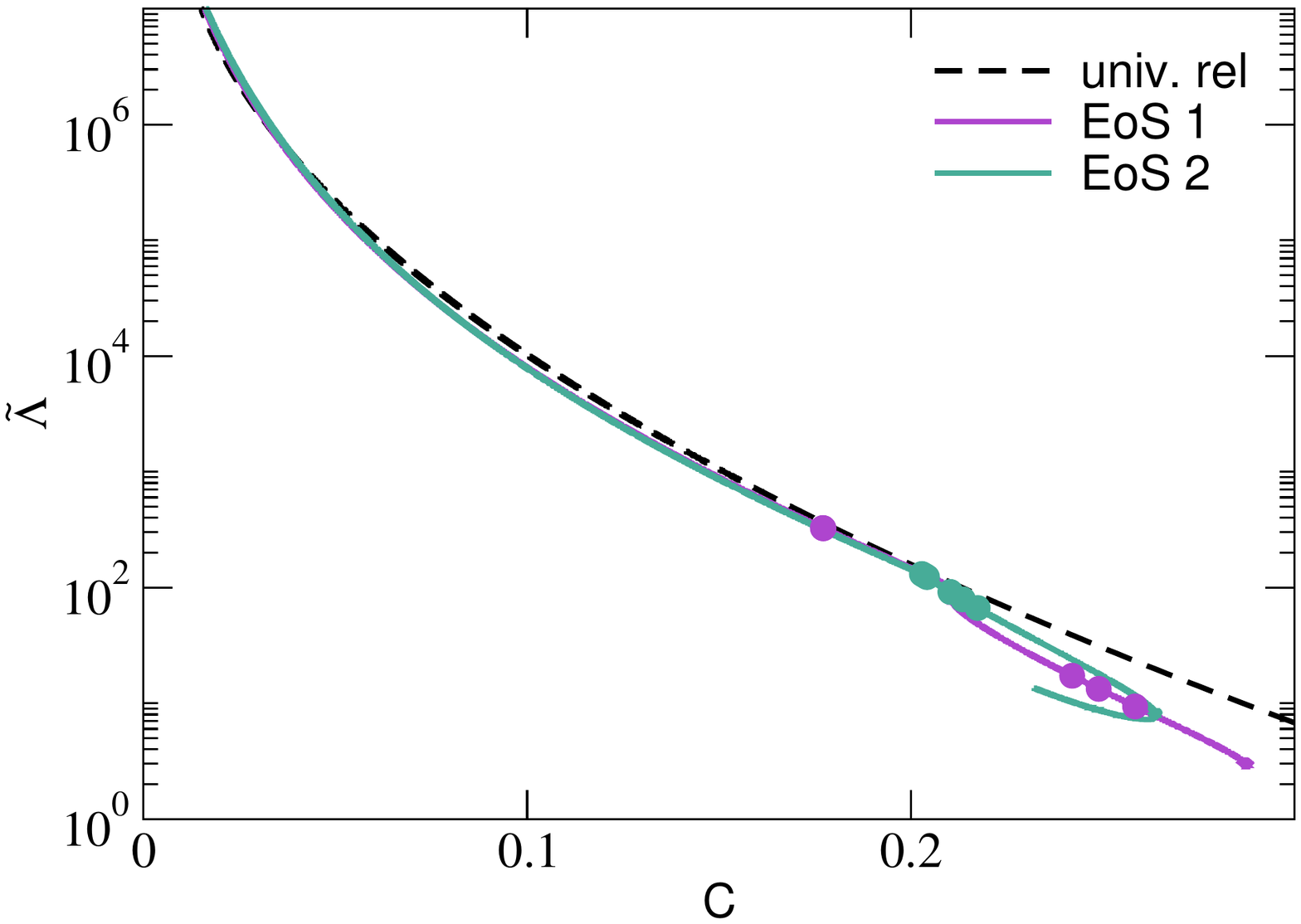}
\caption{Top panel: tidal deformability vs. stellar mass for both equations of state. Bottom panel: tidal deformability vs. compactness $M/R$ for both equations of state. The dashed line shows the universal relation fit extracted from Ref.~\cite{Maselli:2013mva}
\label{tidal}}
\end{figure}

As can be seen in the bottom panel of Fig.~\ref{twin_ver}, EoS 1 has a pretty sparse region in radius between stable twin stars but, for the EoS 2, this difference in radius is much smaller. This fact alone is already enough to infer that the twins in EoS 1 will differ more in their thermal evolution than in the case of the twins in EoS 2. The same idea can be applied to the central stellar pressure implying that the differences generated by the change in composition of matter on the twins of the EoS 1 are much larger than for the twins in the EoS 2. For properties of twin stars pairs generated in each EoS, see Tab.~\ref{tabela}. 

We also calculate the dimensionless tidal deformability for both EoSs following the expression from Ref.~\cite{Favata:2013rwa}. We plot the tidal deformability as a function of stellar masses (top panel) and as a function of compactness $R/M$ (bottom panel) in Fig.~\ref{tidal}. It can be seen in the top panel of the figure how the compact branch of the twin stars (specially in EoS 1) has a much smaller tidal deformability than the less compact one. In the bottom panel, it can be seen how the more compact twin of EoS 1 deviates from the lines formed by its original branch and both branches of EoS 2. It can also be seen how all smaller more compact twin stars deviate from the universal relation fit for these quantities also shown in the figure. The fit was extracted from Ref.~\cite{Maselli:2013mva} and calculated using several hadronic EoSs. See Ref.~\cite{Yagi:2016bkt} for a  comprehensive review on universal relations and Refs.~\cite{Bandyopadhyay:2017dvi,Paschalidis:2017qmb,Gomes:2018eiv,Alvarez-Castillo:2018pve,Sieniawska:2018zzj,Christian:2018jyd,Blaschke:2019tbh,
Li:2019fqe,Christian:2019qer,Benitez:2020fup,Tan:2021nat} for discussions of universal relations in the context of twin stars.

\section{Thermal Evolution}

\subsection{Cooling of Twins}

In this section, we focus on the thermal evolution of twin stars, devoting special attention to the 
difference in thermal behavior among the twin pairs presented in the previous sections. Our goals is to identify whether these stars exhibit different thermal behavior, despite possessing the same gravitational mass. Here, we qualify the difference in cooling and its origins, as well as investigate if such behavior is shared among different microscopic models exhibiting twin stars configuration.

We recall that the cooling of compact stars is governed by the emission of neutrinos and photons from the stellar surface. The equations that describe the thermal evolution for a spherically symmetric, relativistic star are given by \cite{2006NuPhA.777..497P,1999Weber..book,1996NuPhA.605..531S}
\begin{eqnarray}
  \frac{ \partial (l e^{2\phi})}{\partial m}& = 
  &-\frac{1}{\rho \sqrt{1 - 2m/r}} \left( \epsilon_\nu 
    e^{2\phi} + c_v \frac{\partial (T e^\phi) }{\partial t} \right) \, , 
  \label{coeq1}  \\
  \frac{\partial (T e^\phi)}{\partial m} &=& - 
  \frac{(l e^{\phi})}{16 \pi^2 r^4 \kappa \rho \sqrt{1 - 2m/r}} 
  \label{coeq2} 
  \, ,
\end{eqnarray}
where the variables $r$, $\rho(r)$ and $m(r)$, represent the radial distance, the
energy density, and the stellar mass, respectively. The thermal properties are represented by the temperature $T(r,t)$, 
luminosity $l(r,t)$, neutrino emissivity $\epsilon_\nu(r,T)$, thermal conductivity
$\kappa(r,T)$ and specific heat $c_v(r,T)$. 
We also set appropriate boundary conditions at the center of the star (where the heat flow vanishes) 
and at the surface \cite{Gudmundsson1982,Gudmundsson1983,Page2006}. 

In this study, we consider the widely  accepted 
neutrino emission processes that may take place in compact stars, namely: direct Urca process (DU), 
modified Urca process (MU) and Bremsstrahlung (BR) process (for the stellar core), whereas in the crust we consider plasmon decay, in addition to the Bremsstrahlung process. Analogous quark processes are also taken into account wherever appropriate. We refer the reader to references \cite{Yakovlev2000,Yakovlev2004,Page2004} for a detailed review of the neutrino  emission processes in the cooling of neutron stars.

Initially we do not consider any sort of pairing, neither for the hadronic nor for the quark phases. We note that this is not realistic, as most results seem to indicate that observed neutron-star thermal data is most 
likely described for objects with pairing to some level. Nonetheless, we begin our analysis this way to properly 
 probe qualitative differences between the cooling of twin stars pairs. Pairing effects will be discussed in the following subsection.

\begin{figure}
\vspace{3mm}
\includegraphics[width=8.8cm,clip,trim=0cm 0 0 .5cm]{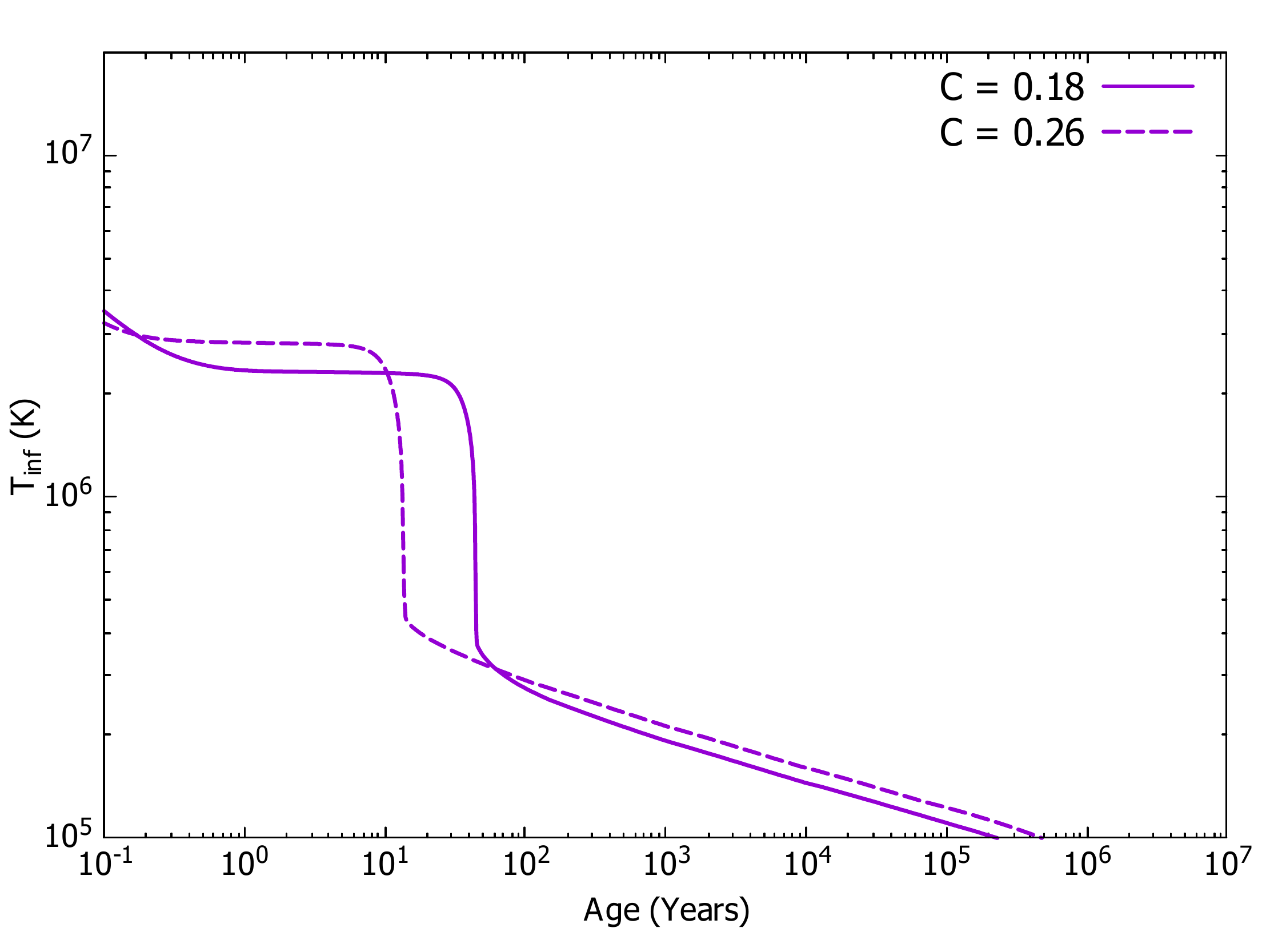}
\caption{Thermal evolution for the most massive twin stars reproduced by EoS 1.
\label{coolver}}
\end{figure}

We begin our analysis by showing the thermal evolution of the most massive twin pair of EoS 1  (whose properties are displayed in Tab.~\ref{tabela}). The cooling of these stars are shown in Fig.~\ref{coolver} - where each star is represented by their respective compactness, $C$. This result shows a prominent difference in the thermal relaxation time of each star (characterized by the sudden drop in surface temperature \cite{Lattimer1994,Gnedin2001}) with the less compact object ($C = 0.18$) thermalizing about 40 years later than their more compact counter-part. The difference in observed surface temperature at older ages, when the star interior is already isothermal, is explained by the gravitational redshift, which is modulated by surface gravity. In this case, the less compact star has a slightly cooler surface temperature, which can be understood by their significantly larger radius and, thus, lower surface gravity. 

\begin{figure}
\vspace{3mm}
\includegraphics[width=8.8cm,clip,trim=0cm 0 0 .5cm]{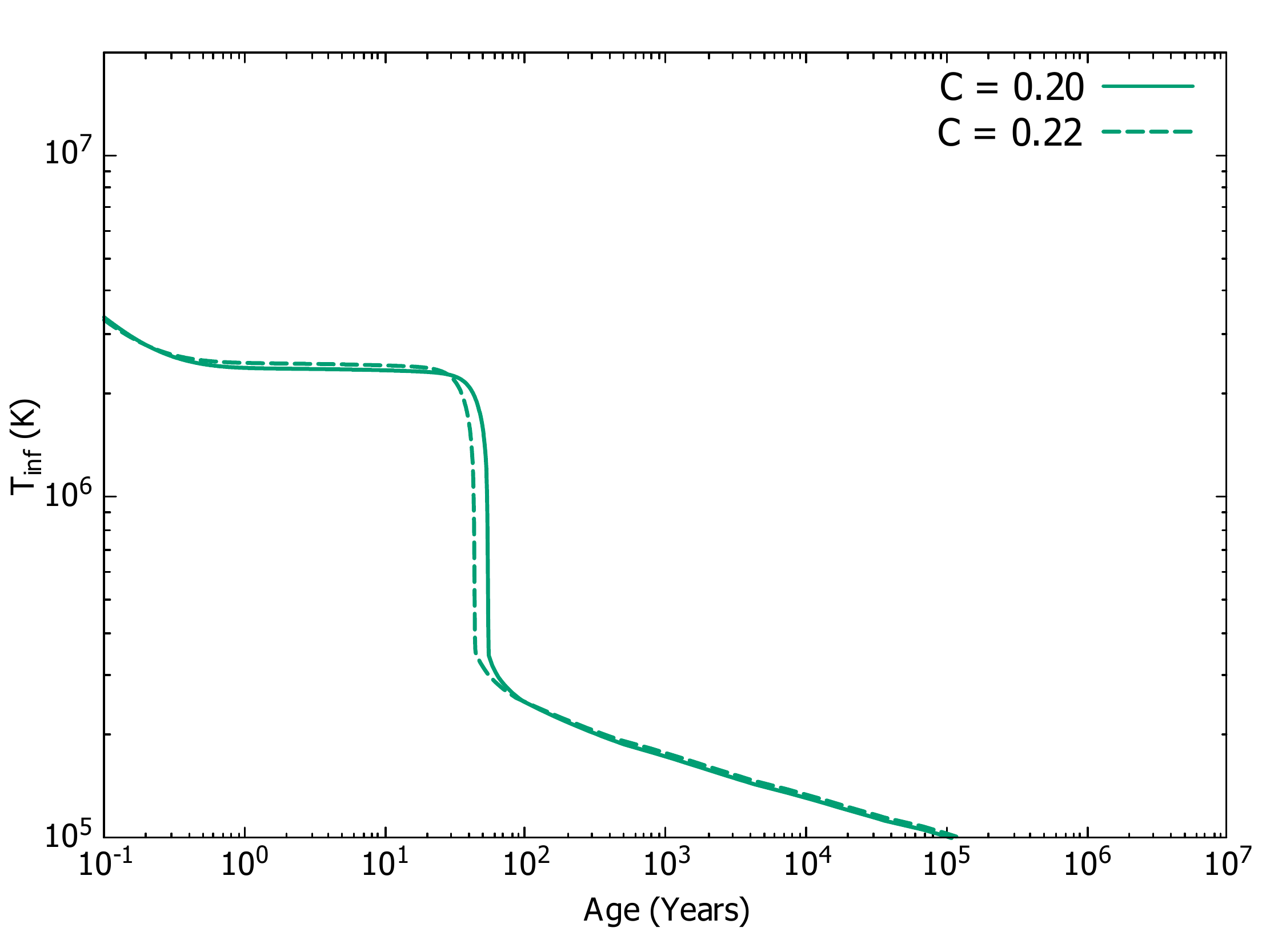}
\caption{Same as Fig.~\ref{coolver} but for EoS 2. 
\label{coolros}}
\end{figure}

As for the twin pair of EoS 2 (see once more table ~\ref{tabela}), their thermal evolution is shown in Fig.~\ref{coolros}. For this EoS the difference in cooling is less pronounced, with both stars exhibiting a 
very similar thermal behavior, and only a $\sim 10$ years delay on their thermal relaxation time. Nevertheless, the trends are the same as the ones discussed for EoS 1.


These results show that for the models studied, qualitatively, there is not much difference in the thermal behavior shown by the twin pairs. Both stars exhibit fast cooling, as indicated by the large temperature drop at the thermal relaxation age -- this can be understood by the high density of these stars that guarantees the DU process to be present, thus leading to a fast cooling scenario. There is, however, a quantitative difference in the thermal relaxation time of the twins. This can be understood by analyzing  the compactness of each star. For the twins of EoS 1 we see a relatively large discrepancy in their compactness, with the most compact twin having $C = 0.26$ while the less compact has $C= 0.18$. On another hand, the twins of EoS 2 have a much more similar compactness ($C=0.22$ and $C=0.20$ respectively).
 This result seems to be in line with the analysis of the thermal relaxation of neutron stars done in \cite{Lattimer1994,Gnedin2001,2020A&A...642A..42S}, which have found that the thermal relaxation of a neutron star can be well described by 

\begin{equation}
t_w = t_1 \times \left(\frac{\Delta R_{C}}{1 km}\right)^2 \left( 1- 2 C\right)^{-3/2}.
\label{eq:rel_time}
\end{equation}

One should note that in the studies mentioned above, twin-star configurations were not considered. 
Their results show, however, that the relaxation time mostly depends of macroscopic properties, namely the mass ($M$), radius ($R$) and crust thickness ($\Delta R_{C}$). The microscopic information is carried by the constant $t_1$ which varies according to the microscopic model. 
Based on this, we believe that regardless of the model, the difference in relaxation time for twin stars depends mostly on the difference in their compactness, rather than the difference in their inner composition. However, this should be regarded with care, as we have only studied two models that allow for twin stars. We intend to test in the future more such models (as they become publicly available) in order to perform a systematic study to infer if such behavior is general. 

\subsection{Pairing Effects}

\begin{figure}
\vspace{3mm}
\includegraphics[width=10.cm,clip,trim=1.87cm .5cm 0 1.5cm]{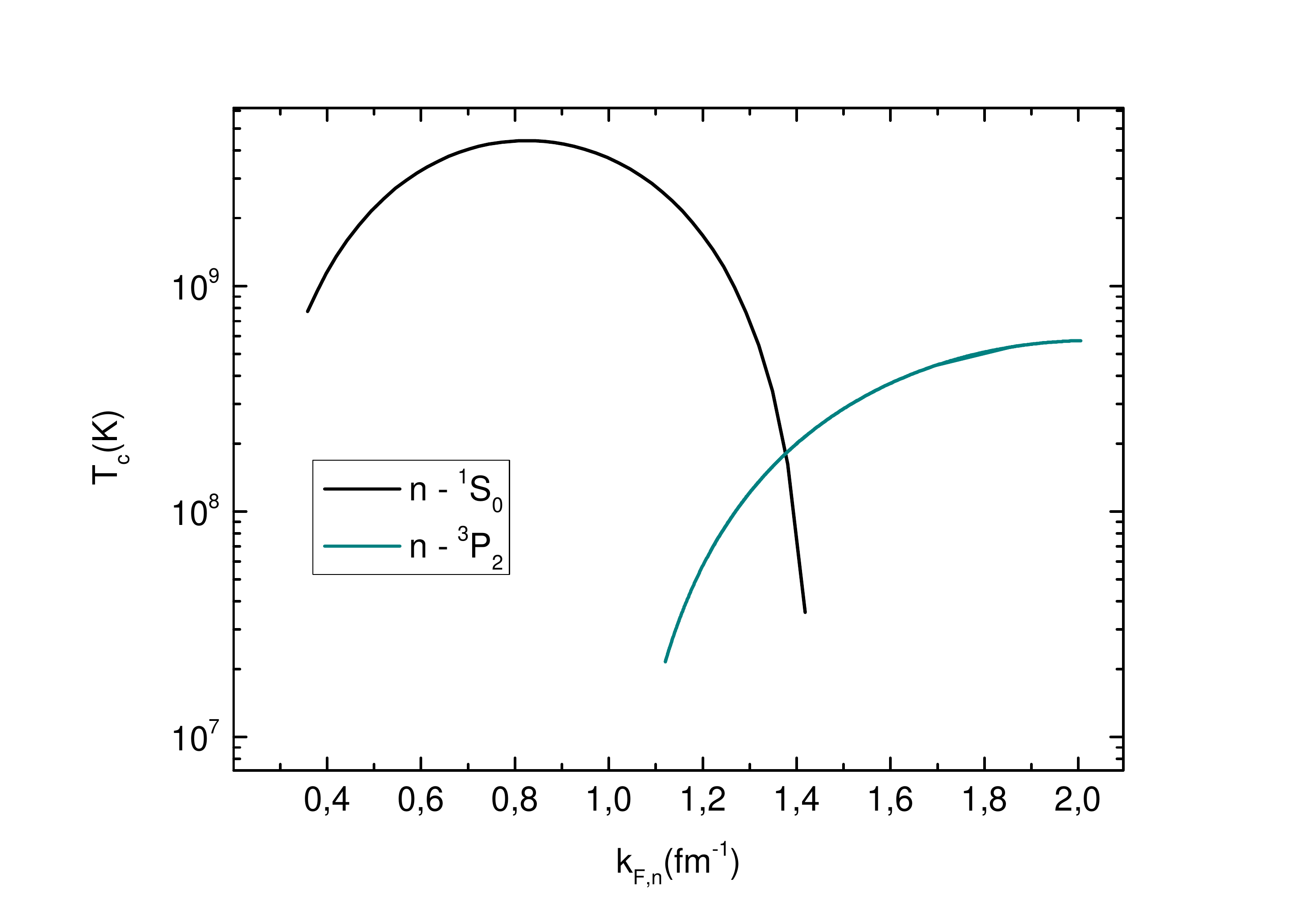}
\caption{Critical temperature as a function of neutron Fermi momentum for both triplet and singlet neutron pairs.
\label{SFnn}}
\end{figure}

Having studied the general aspects of the cooling of twin stars, we now devote our attention to more realistic 
cooling behavior, taking into account pairing in both phases. 
We recall that the thermal evolution of stars under EOS 1 was thoroughly studied in 
ref. \cite{PhysRevC.91.055808}, and here we are focusing only in the cooling of twin pairs, 
and their potential differences. 
For the hadrons, we consider neutron-neutron pairs that may form a singlet or a triplet. As discussed in Refs.~\cite{Page2004,Beloin:2016zop}, neutron singlet pairing takes place in low density regime, mostly in the stellar crust, whereas triplet may extend into the core. We have applied a pairing model previously used to model the cooling of neutron stars \cite{Negreiros_2018}. We show in Fig.~\ref{SFnn} the critical temperature for the different neutron pairing patterns, used across the different EoS's studied.

\begin{figure}
\vspace{3mm}
\includegraphics[width=10.cm,clip,trim=1.87cm .5cm 0 1.5cm]{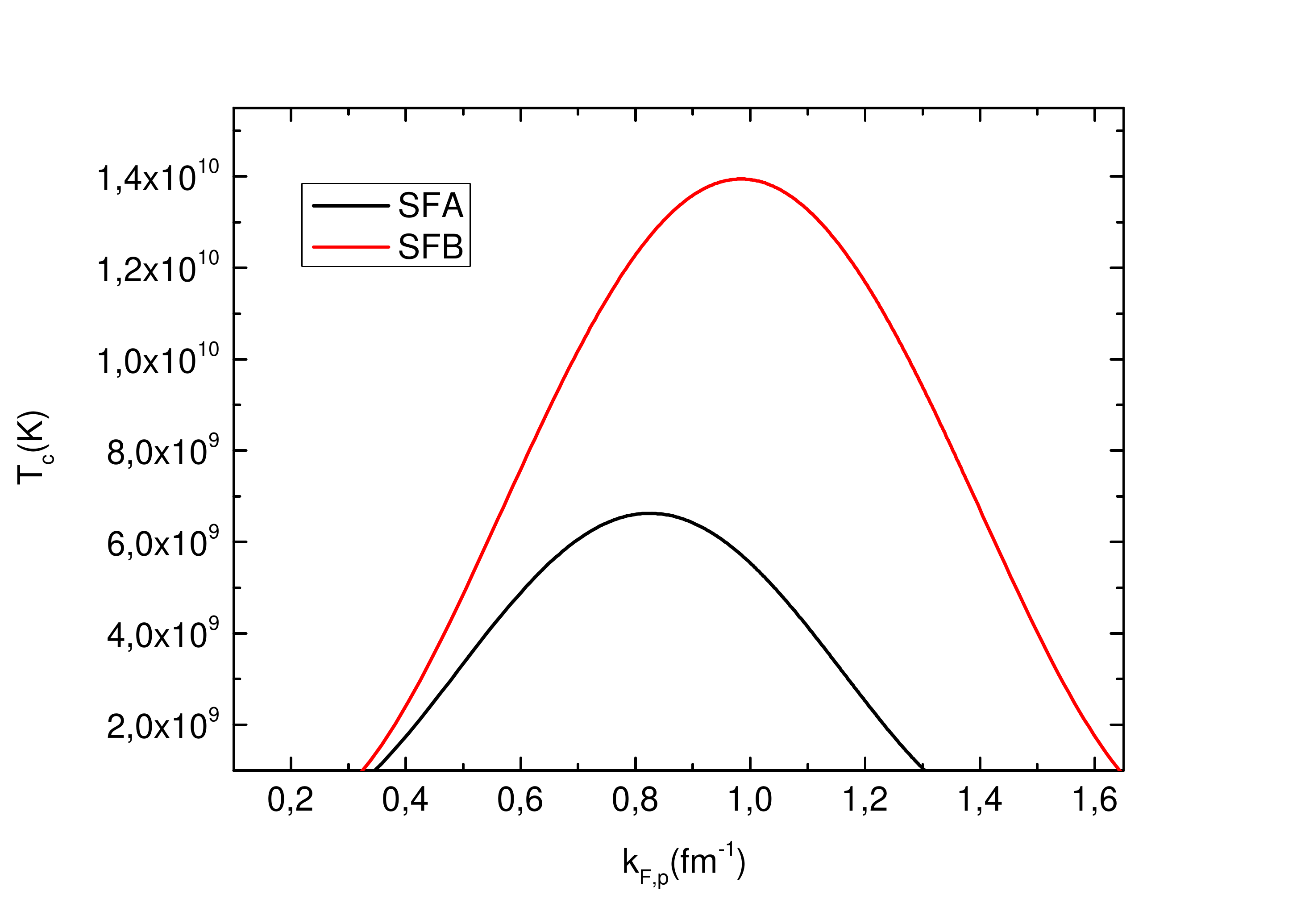}
\caption{Critical temperature as a function of the proton Fermi momentum for $pp$ singlet pairs in both proton pairing models considered.
\label{SFp}}
\end{figure}

Regarding proton pairing, the current picture is less clear. It is generally accepted that protons may form singlet pairs in the core of neutron stars, however, the extent and magnitude of the pairing gaps is still subject to much debate. Most of observed data seem to indicate that some level of proton pairing is necessary, otherwise most neutron stars would cool down much too quickly \cite{Page2011a,Page2009,Shternin2011a}. To remedy some of the uncertainty with regards to proton pairing, we explore two pairing models, covering a moderate (SFA) and a more extensive (SFB) proton pairing. The critical temperature as a function of the proton's Fermi momentum is shown in Fig.~\ref{SFp}. Note that the presence of hadronic pairing leads to the pair-breaking-formation process (PBF) \cite{Yakovlev2000}, which is a transient neutrino emission mechanism that takes place near the onset of hadronic pairing. Such process is appropriately taken into account in our calculations.

Finally, we also consider the possibility of quark pairing. We allow quarks to be in a Color-Flavor-Locked (CFL) \cite{Alford2005,Alford2005a,RevModPhys.80.1455} state and study the possibility of pairing gaps of $\Delta = 10$ MeV. Note that pairing of this magnitude should not affect the EoS in any appreciable way \cite{PhysRevD.67.074024}. Due to the quark pairing uncertainties, we have considered the prescription of Ref.~\cite{Blaschke2000a}, in which the CFL critical temperature is given by $T_c = 0.4~\Delta$. Quark direct Urca processes are thus suppressed by a factor $e^{-\Delta/T}$, whereas the modified Urca and Bremsstrahlung processes by $e^{-2\Delta/T}$. Finally, the quark specific heat is modified by a factor $f$ given by \cite{Blaschke2000a}
\begin{equation}
f = 3.2\left(\frac{T_c}{T}\right)\left(2.5 - 1.7\frac{T}{T_c}+3.6\left(\frac{T}{T_c}\right)^2\right)e^{-\Delta/T}.
\label{eq:Qcvf}
\end{equation}

\begin{figure}
\vspace{3mm}
\includegraphics[width=10.cm,clip,trim=1.7cm 0 0 .5cm]{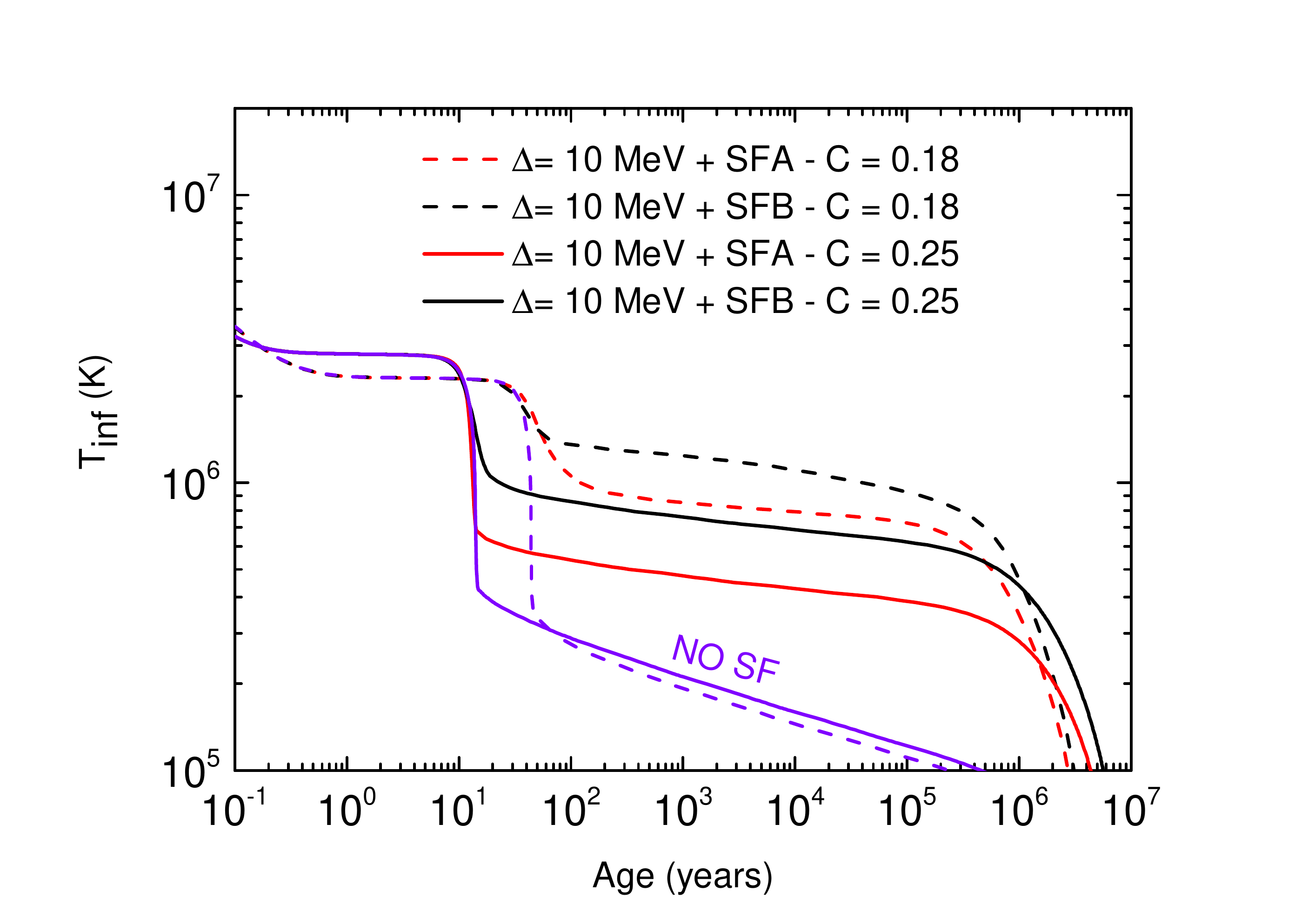}
\caption{Thermal evolution of the most massive twins of EoS 1. Each twin is identified by its compactness. We show both proton pairing models used, identified by SFA and SFB. All curves consider the presence of CFL pairing for quark matter with gap $\Delta = 10$ MeV. Also shown is the previously studied case (Fig.~\ref{coolver}) with no pairing, indicated by the NO SF label.
\label{EOS1_SF}}
\end{figure}

The thermal evolution of the twins in the EoS 1 model, with pairing taken into account, can be found in Fig.~\ref{EOS1_SF}. Our results indicate that the presence of pairing not only slows down the cooling of all stars, but also makes the difference in the thermal behavior among twins pairs more distinguishable. For both proton pairing models considered, the less compact twin cools down more slowly, which is consistent with suppression effects that pairing has on the neutrino emission processes - as pairing drastically reduces the neutrino emission from the DU process. This leads to an interesting phenomenon of two stars with the same mass exhibiting largely distinct thermal evolution. This phenomenon can only be observed in twin stars - and to the extent of our knowledge has not been reported before. We also note that the thermal relaxation age is not modified (since pairing does not change macroscopic properties of the stars) - only the slope of the cooling curve at the thermal relaxation changes. This is consistent with thermal relaxation theory, as superfluidity is expected to change the parameter $t_1$ for Eq.~(\ref{eq:rel_time}) in a microscopic model.

\begin{figure}
\vspace{3mm}
\includegraphics[width=10.cm,clip,trim=1.7cm 0 0 .5cm]{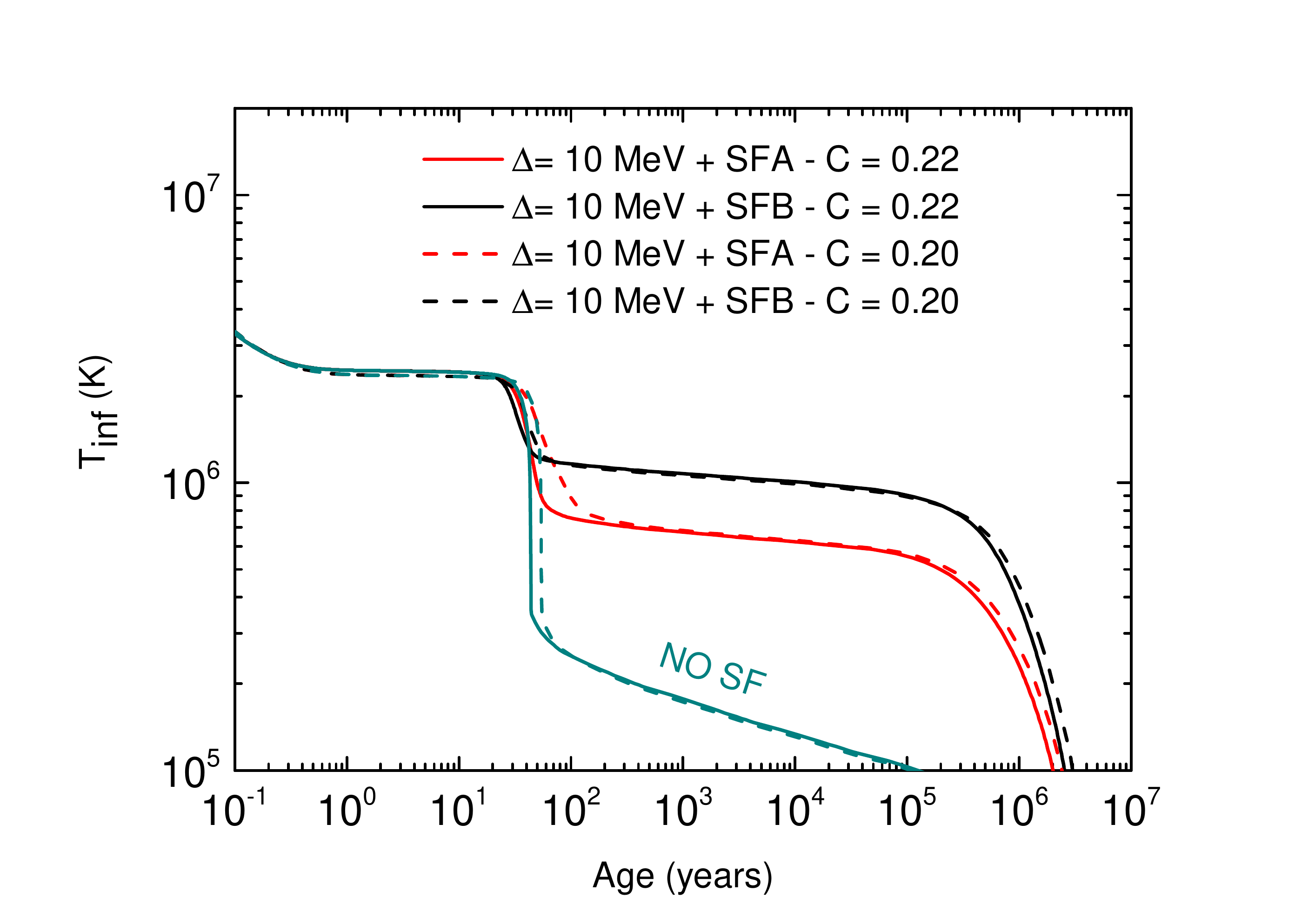}
\caption{Same as Fig.~\ref{EOS1_SF} but for equation of state 2.}
\label{EOS2_SF}
\end{figure}

We now show the results for the twins of EoS 2 under the effects of pairing in Fig.~\ref{EOS2_SF}. As expected, once more the presence of pairing leads to a substantial slow down of the cooling in all stars. For this model, however, the presence of pairing does not affect the cooling of one twin star in the pair more prominently. This is due to the fact that this particular model exhibits twins with similar (and relatively high compactness) - thus subjecting them both to the same consequences of pairing.

\section{Conclusions}

We have performed an in depth analysis of the cooling of twin pairs for two different available equation of state (EoS) models. Note that for thermal evolution studies one needs to know the particle population, thus only EoSs that are generated by microscopic models can be used, which excludes parametrized EoSs (of which consist most of the descriptions that reproduce twin stars). We have used in this work two different microscopic models with substantially different physical motivations and compositions. As a consequence, each model leads to a twin configuration with very different properties. While EoS model 1 has twin stars with masses $\sim 1.67 M_\odot$, the second model studied has much more massive twins $\sim 2.0 M_\odot$ (although both models can reproduce $\sim 2.0 M_\odot$ stars). Furthermore, due to the different nature of the microscopic models studied, the two sets of pairs cover very different ranges of central densities, with the twins of EoS 1 having central number densities of 0.4 -- 1.5 $fm^{-3}$ and the twins of EoS 2 having 0.6 -- 0.7 $fm^{-3}$. This behavior is reflected in compactness of each twin set, with the twins of EoS 1 having a large difference in compactness, while those of EoS 2 being much more similar. 

In order to investigate whether such characteristics have any effect on the stellar thermal behavior, we have performed a thorough investigation of the cooling of the most massive twin pair for each of the models investigated. Our first analysis consisted of a simple study of the thermal evolution of each pair, taking into account only the thermal processes allowed by each microscopic composition. This study showed us that the most distinguishable difference between each twin for each set is the thermal relaxation time. Such difference is most evident for the twins of EoS 1, with thermal relaxation taking place 40 years a part. The twins of EoS 2 have a much more similar behavior, with thermal relaxation only 10 years apart. We conclude that such differences are associated with the macroscopic properties of such stars, particularly their compactness. The twins of EoS 1 that have substantially different compactnesses present a most prominent discrepancy in the thermal relaxation, whereas the twins of EoS 2, that have similar compactness, exhibit a much more similar thermal behavior. This results are in agreement with the general properties of thermal relaxation in neutron stars, as studied in Refs.~\cite{Lattimer1994,Gnedin2001,2020A&A...642A..42S}.

With the goal of performing a complete study of the cooling of twin stars, we have also investigated how pairing affects the thermal evolution. For that, we have included neutron, proton, and quark pairing. We considered the possibility of neutron singlet and triplet pairing, covering regions respective to the stellar crust, as well as the core. For proton pairing we have considered singlet pairs. Due to the current uncertainties regarding the extension and magnitude of proton pairing, we have opted to consider two proton pairing scenarios: a moderate and a more pervasive one. For quark pairing, we considered the possibility of CFL. 

Our results show that, as expected, pairing significantly slows down the cooling in both models. Differently than the case without pairing, however, the cooling of the twins of EoS 1 exhibited significant difference, beyond that of the thermal relaxation age (which is still present). Due to the large difference in compactness between each twin of EoS 1, pairing has different effects on each twin, having less of an impact on the most compact. As for EoS 2, in which the twins are much more similar, pairing affects them mostly similarly, thus leading to no additional noticeable difference in the thermal behavior among the twins. We believe that this is due to the fact that in this model the twins have a very similar compactness; As for the impact of different scenarios of proton pairing, not surprisingly, the most pervasive proton pairing leads to slower cooling in all cases studied.

Our results allow us to draw a few conclusions. The first is that this study is in agreement with previous analysis of the thermal relaxation of neutron stars, in which the thermal relaxation age seems to be directly connected to macroscopic properties \cite{Lattimer1994,Gnedin2001,2020A&A...642A..42S}, namely the mass and radii (represented by the compactness), even though the original study did not consider the possibility of twin stars, our study seems to indicate that such results are still valid in this case. Our investigation shows that other than the thermal relaxation age, the cooling of the twins studied are qualitatively very similar. We have noted that two conditions need to be met for the twins to exhibit significantly different thermal evolutions: 1) substantial difference in compactness, and 2) pairing. The combination of these two phenomena may lead to the interesting phenomena of two stars with the same mass to exhibit qualitatively different thermal behavior. This can be understood as a novel way to study stellar compactness, quark deconfinement, and phase transitions, even in extremely magnetized and/or isolated stars, in which case techniques used for observations of NICER and LIGO/VIRGO could not be applied. Evidently, our study is limited to currently available twin-star models, thus we cannot generalize this assessment to all twin stars, although it seems reasonable to believe that this will always be the case considering how distinct the models studied were.

\section*{Acknowledgements}

F.L. and L.M acknowledges financial support from CAPES. R.N.\ acknowledges financial support from CAPES, CNPq, and FAPERJ. This work is part of the project INCT-FNA Proc. No. 464898/2014-5 as well as FAPERJ JCNE Proc. No. E-26/203.299/2017. 
Support for this research comes from the National Science Foundation under grants PHY1748621, NP3M PHY-2116686, and MUSES OAC-2103680, and from  PHAROS (COST Action CA16214).

\bibliographystyle{apsrev4-1}
\bibliography{references}
\end{document}